# In-Out impurity density asymmetry due to the Coriolis force in a rotating tokamak plasma

Chengkang Pan[1], Shaojie Wang[2], Xiaotao Xiao[1], Lei Ye[1], Yingfeng Xu[1] and Zongliang Dai[2]

[1]*Institute of Plasma Physics, Chinese Academy of Sciences, Hefei 230031, People's Republic of China*

[2]*Department of Modern Physics, University of Science and Technology of China, Hefei 230026, China*

E-mail: ckpan@ipp.ac.cn

**Abstract**

The effect of the Coriolis force due to the impurity toroidal and poloidal rotation on the in-out impurity density asymmetry in a rotating tokamak plasma is identified. The in-out impurity density asymmetry can be induced by the Coriolis force with $q \mathrm{v}_{\theta,z} \omega_z > 0$, in this case, when moving along the magnetic field line from the outboard side to the inboard side in a magnetic flux surface, one sees a positive Coriolis force. The proposed theory is consistent with the ASDEX Upgrade experimental observations.

1. Introduction

In the magnetic confinement fusion device, such as a tokamak, the heavy impurity ions exist due to the inward radial transport of the impurities released from the divertor plate and the first wall. Impurity transport, especially impurity pinch [1-3], is one of the hot topics in the magnetic confinement fusion area. Impurity accumulation in the core of a tokamak plasma is well known to be a big challenge because of fuel dilution and power loss from radiation. This has been an issue of great concern for several decades. Reduction or suppression of the core impurity accumulation is essential to realize a magnetic confinement fusion reactor, such as the International Thermonuclear Experimental Reactor (ITER) [4]. The experimental observations in the tokamaks [5-9] strongly support that the core impurity accumulation is mainly driven by the neoclassical inward convection [10]. Both the theoretical predictions [11-14] and experimental observations [15] show that the poloidally asymmetric impurity density distribution on a magnetic flux surface can significantly affect the impurity neoclassical transport; the impurity accumulation on the inboard side of a flux surface (in-out asymmetry) can reverse the neoclassical convection from inward to favorable outward direction [13, 16], which is beneficial for removing the impurities from the plasma core. Therefore it is worthwhile to understand the physical mechanisms that drive the in-out impurity asymmetry to avoid the impurity core accumulation through the external controlling.

The well-known centrifugal force effect due to the impurity toroidal rotation in a tokamak plasma can push the impurities to accumulate on the outboard side of the flux surface (out-in asymmetry), which has been pointed out theoretically by Hinton, Wong [17] and Wesson [18] earlier and observed in the tokamak experiments [19, 20]. But the centrifugal force effect can not explain the in-out impurity asymmetry observed in JET [21, 22], Alcator C-Mod [23-28], ASDEX Upgrade [29, 30] tokamaks. To understand the in-out impurity asymmetry, the impurity-ion friction force model [31-33] and the ion cyclotron resonance heating (ICRH)-induced poloidal electric field model [25, 34] were proposed. The impurity-ion friction model was applied to explain the in-out asymmetry of the heavy impurities in the edge of the Alcator C-Mod tokamak [24, 26]; the in-out asymmetry of the heavy impurities observed in the core plasma in the JET [21, 22] and Alcator C-Mod [24, 25] tokamak can be due to the ICRH-induced poloidal electric field; the in-out asymmetry of the light impurities observed in the edge of the ASDEX Upgrade [29, 30] was thought to be mainly due to the combining effect of the impurity-ion friction and the poloidal centrifugal force. The experimental observations in the Alcator C-Mod [23] and ASDEX Upgrade [29, 30] tokamak show that the impurity poloidal rotation is strongly related to the in-out impurity density asymmetry. It will be shown in this Letter that in a rotating tokamak plasma the impurity poloidal rotation combining with the impurity toroidal rotation generates the Coriolis force [35, 36], which can affect the in/out impurity density asymmetries largely and is not included in the previous models [25, 31-34].

The remaining part of this paper is organized as follows. In section 2, the theoretical model will be presented. In section 3, a numerical example will be shown as an application of the proposed theoretical model. In section 4, the summary and conclusion will be presented.

2. Theoretical model

The physical picture of the effect of the Coriolis force on the in/out impurity density asymmetries will be explained firstly. In the toroidally rotating frame in a tokamak plasma, the impurity poloidal rotation generates the Coriolis force, $-2n_z m_z \boldsymbol{\omega}_z \times \mathbf{v}_{\theta,z}$, with $n_z$, $m_z$ the impurity density and particle mass, respectively; here $\boldsymbol{\omega}_z$ and $\mathbf{v}_{\theta,z}$ are the toroidal rotation angular velocity and the poloidal rotation velocity of the impurities, respectively. This Coriolis force is in the toroidal direction, which can push the impurities along the magnetic field line. When moving along the magnetic field line from the outboard side to the inboard side in a flux surface, if the impurity sees a positive Coriolis force, the impurity is pushed by the Coriolis force to the inboard side and the in-out impurity asymmetry can be induced, otherwise the impurity is pushed to the outboard side and the out-in impurity asymmetry due to the centrifugal force is enhanced further. Therefore the effect of the Coriolis force on the in/out impurity asymmetries depends on both the direction of the Coriolis force determined by $-2n_z m_z \boldsymbol{\omega}_z \times \mathbf{v}_{\theta,z}$ and the helicity of the magnetic field line from the outboard side to the inboard side determined by the sign of $q = rB_\varphi / R_0 B_\theta$, whose absolute value is the usual safety factor in the tokamaks of large aspect-ratio with circular cross-section. Here $r$, $R_0$ are the minor radius of the flux surface and the major radius at the magnetic axis, respectively; $B_\varphi$, $B_\theta$ are the toroidal and poloidal components of the magnetic field. This new mechanism, which has not been pointed out in the previous studies [25, 31-34], is schematically illustrated in Fig. 1.

To proceed, we begin with the steady-state equations of continuity and momentum,

$$\nabla \cdot (n_z \mathbf{v}_z) = 0, \tag{1}$$

$$n_z m_z \mathbf{v}_z \cdot \nabla \mathbf{v}_z = -\nabla p_z + Z e n_z \mathbf{v}_z \times \boldsymbol{B} - Z e n_z \nabla \Phi, \tag{2}$$

with $\mathbf{v}_z$ the impurity fluid velocity, $p_z$ the fluid pressure, $Ze$ the charge of the impurity ion; $\boldsymbol{B}$ is the magnetic field. Here we will focus on the inertial effect on the in/out impurity density asymmetries and ignore the impurity-ion friction force, which has been widely discussed in [24, 26 30-33].

The magnetic field is given by $\boldsymbol{B} = I \nabla \varphi + \nabla \varphi \times \nabla \psi \equiv B_\varphi \hat{\boldsymbol{e}}_\varphi + B_\theta \hat{\boldsymbol{e}}_\theta$. We choose a right-hand coordinate system $(\psi, \theta, \varphi)$.

The lowest order divergence-free impurity mass flow, which satisfies Eq. (1), can be written as [37-40]

$$n_z \mathbf{v}_z = \nabla \varphi \times \nabla \Theta_z + n_z \mathrm{v}_{\varphi,z} \hat{\boldsymbol{e}}_\varphi, \tag{3}$$

Where $\Theta_z$ is a function describing the impurity poloidal rotation $v_{\theta,z} = \dfrac{1}{J|\nabla\theta|}\dfrac{\partial\Theta_z}{\partial\psi}$

and the impurity flow normal to the flux surface $v_{\psi,z} = -\dfrac{1}{J|\nabla\psi|}\dfrac{\partial\Theta_z}{\partial\theta}$ and

$J = (\nabla\psi\cdot\nabla\theta\times\nabla\varphi)^{-1}$.

Substituting Eq. (3) into Eq. (2), one can write the $\theta$- and $\varphi$-components of the inertia term in Eq. (2) as

$$\hat{e}_\theta\cdot(n_z m_z \mathbf{v}_z\cdot\nabla\mathbf{v}_z) = -\dfrac{n_z m_z v_{\varphi,z}^2}{J|\nabla\psi|}\dfrac{\partial R}{\partial\theta} + |\nabla\theta||\nabla\psi|n_z m_z v_{\psi,z}\dfrac{\partial}{\partial\psi}(\dfrac{v_{\theta,z}}{|\nabla\theta|})$$
$$+|\nabla\theta|n_z m_z v_{\theta,z}\dfrac{\partial v_{\theta,z}}{\partial\theta} - |\nabla\theta||\nabla\psi|n_z m_z v_{\psi,z}^2\dfrac{\partial}{\partial\theta}(\dfrac{1}{|\nabla\psi|}) \qquad (4)$$

$$\hat{e}_\varphi\cdot(n_z m_z \mathbf{v}_z\cdot\nabla\mathbf{v}_z) = \dfrac{n_z m_z v_{\theta,z}}{J|\nabla\psi|}\dfrac{\partial(R v_{\varphi,z})}{\partial\theta} + \dfrac{n_z m_z v_{\psi,z}}{J|\nabla\theta|}\dfrac{\partial(R v_{\varphi,z})}{\partial\psi}. \qquad (5)$$

From the $\varphi$-component of Eq. (2), one finds

$\partial(\Theta_z, R v_{\varphi,z} - Ze\psi/m_z)/\partial(\psi,\theta) = 0$, which yields $\Theta_z = \Theta_z(R v_{\varphi,z} - Ze\psi/m_z)$. It implies that there is non-zero impurity flow normal to the flux surface to ensure the equilibrium force balance in a rotating tokamak plasma [37-40]. Actually one can obtain $v_{\psi,z} = v_{\theta,z}\dfrac{|\nabla\theta|}{R}\dfrac{\partial(R v_{\varphi,z})}{\partial\theta}(\Omega_{\theta,z} - \dfrac{|\nabla\psi|}{R}\dfrac{\partial(R v_{\varphi,z})}{\partial\psi})^{-1}$. Here $\Omega_{\theta,z}$ is the impurity ion gyro-frequency evaluated with the poloidal magnetic field. With the tokamak plasma of circular cross-section and the impurity toroidal rotation assumed to be rigid on the flux surface, the impurity flow normal to the flux surface is estimated to be $v_{z,\psi} \approx -2 v_{\theta,z}(\omega_z/\Omega_{\theta,z})\sin\theta \ll v_{\theta,z} \leq v_{\varphi,z}$ if $\omega_z \ll \Omega_{\theta,z}$, which is satisfied with $\omega_z \leq 10^6 krad/s$ and $\Omega_{\theta,z} \sim 10^7 s^{-1}$ in the typical tokamak plasmas. Therefore the effect of radial flow, $v_{z,\psi}$, shall be ignored in the following. Then Eqs. (4) and (5) are

reduced to

$$\hat{e}_\theta \cdot (n_z m_z \mathbf{v}_z \cdot \nabla \mathbf{v}_z) = -\frac{n_z m_z v_{\varphi,z}^2}{J|\nabla\psi|}\frac{\partial R}{\partial \theta} + |\nabla\theta| n_z m_z v_{\theta,z}\frac{\partial v_{\theta,z}}{\partial \theta}, \tag{6}$$

$$\hat{e}_\varphi \cdot (n_z m_z \mathbf{v}_z \cdot \nabla \mathbf{v}_z) = \frac{n_z m_z v_{\theta,z}}{J|\nabla\psi|}\frac{\partial (R v_{\varphi,z})}{\partial \theta}. \tag{7}$$

The right-hand-side of Eq. (7) corresponds to the Coriolis force in the toroidally rotating frame generated by the impurity poloidal rotation.

The impurity density poloidal variation can be obtained through the $\mathbf{B}$-parallel component of Eq. (2)

$$\mathbf{b}\cdot(n_z m_z \mathbf{v}_z \cdot \nabla \mathbf{v}_z) = -\mathbf{b}\cdot\nabla p_z + \mathbf{b}\cdot Zen_z(\mathbf{v}_z\times\mathbf{B}) - \mathbf{b}\cdot(Zen_z\nabla\Phi), \tag{8a}$$

$$\frac{B_\theta}{B}\hat{e}_\theta\cdot(n_z m_z \mathbf{v}_z \cdot \nabla \mathbf{v}_z) + \frac{B_\varphi}{B}\hat{e}_\varphi\cdot(n_z m_z \mathbf{v}_z \cdot \nabla \mathbf{v}_z) = -\frac{B_\theta}{B}|\nabla\theta|\frac{T_z\partial n_z}{\partial\theta} - Zen_z\frac{B_\theta}{B}|\nabla\theta|\frac{\partial\Phi}{\partial\theta}$$

$$\tag{8b}$$

To obtain Eq. (8b), the impurity temperature is taken as a flux surface function; $\mathbf{b} = \frac{B_\varphi}{B}\hat{e}_\varphi + \frac{B_\theta}{B}\hat{e}_\theta$ and $\mathbf{b}\cdot\nabla = \frac{B_\theta}{B}|\nabla\theta|\frac{\partial}{\partial\theta}$ are used. With the help of Eqs. (6) and (7), it is not hard to derive

$$T_z\frac{\partial n_z/\partial\theta}{n_z} = \frac{m_z v_{\varphi,z}^2}{J|\nabla\psi||\nabla\theta|}\frac{\partial R}{\partial\theta} - \frac{B}{B_\theta}\frac{m_z v_{\theta,z}}{J|\nabla\psi||\nabla\theta|}\frac{\partial(R v_{\varphi,z})}{\partial\theta} - m_z v_{\theta z}\frac{\partial v_{\theta,z}}{\partial\theta} - Ze\frac{\partial\Phi}{\partial\theta}. \tag{9}$$

With $J = (\nabla\psi\cdot\nabla\theta\times\nabla\varphi)^{-1}$, Eq.(9) can also be written as

$$T_z\frac{\partial n_z/\partial\theta}{n_z} = \frac{m_z v_{\varphi,z}^2}{R}\frac{\partial R}{\partial\theta} - \frac{B}{B_\theta}\frac{m_z v_{\theta,z}}{R}\frac{\partial(R v_{\varphi,z})}{\partial\theta} - m_z v_{\theta z}\frac{\partial v_{\theta,z}}{\partial\theta} - Ze\frac{\partial\Phi}{\partial\theta}. \tag{10}$$

The impurity density poloidal variation in a rotating tokamak plasma is determined by Eq. (10), which is valid for the arbitrary aspect-ratio and shaped tokamak plasma. It can be found that the impurity density poloidal variation is affected by (1) the centrifugal force due to the impurity toroidal rotation, (2) the Coriolis force due to the impurity toroidal and poloidal rotation, (3) the poloidal centrifugal force due to the poloidal variation of the impurity poloidal rotation, and (4) the poloidal electric field. The effects of (1) and (4) on the out-in impurity density asymmetry have been discussed in Ref. [17, 18]. The important role of (3) on the in-out impurity density asymmetry has been pointed out in Ref. [29, 30]. The in-out asymmetry factor of the impurity density in a rotating tokamak plasma can be obtained by integrating Eq. (10) directly.

To demonstrate the main underling physics for the in-out impurity density asymmetry, we consider a tokamak plasma of large aspect-ratio with circular cross-section. The impurity toroidal rotation will be assumed to be rigid on every flux surface. Then Eq. (10) will reduce to

$$T_z \frac{\partial n_z / \partial \theta}{n_z} = -m_z \omega_z^2 rR \sin\theta + 2m_z q R_0 \omega_z \mathrm{v}_{\theta,z} \sin\theta - m_z \mathrm{v}_{\theta,z} \frac{\partial \mathrm{v}_{\theta,z}}{\partial \theta} - Ze \frac{\partial \Phi}{\partial \theta}. \qquad (11)$$

To obtain Eq. (11) from Eq. (10), $R = R_0 + r\cos\theta$, $q = rB_\varphi / R_0 B_\theta$ and $B \approx B_\varphi$ are used.

To discuss the respective contributions to the impurity density poloidal variation, the poloidal-dependent impurity poloidal rotation is expanded as $\mathrm{v}_{\theta,z}(\theta) = \overline{\mathrm{v}}_{\theta,z}(1+\delta\cos\theta)$ with the up-down symmetric $\mathrm{v}_{\theta,z}$ considered, where $\overline{\mathrm{v}}_{\theta,z}$ and $\delta\overline{\mathrm{v}}_{\theta,z}$ are the poloidal-independent and the lowest order cosine-component of $\mathrm{v}_{\theta,z}$ respectively. The ratio between the first three terms on the right-hand-side in Eq. (11) will be $1 : (2q/\varepsilon)(\mathrm{v}_{\theta,z}/\mathrm{v}_{\varphi,z}) : (\delta/\varepsilon)(\mathrm{v}_{\theta,z}/\mathrm{v}_{\varphi,z})^2$, where $\varepsilon = r/R_0$. In the core tokamak plasma with $|\mathrm{v}_{\theta,z}/\mathrm{v}_{\varphi,z}| \approx 0$, the effect of the centrifugal force due to the impurity toroidal rotation will be dominant; the effect of the Coriolis force will be comparable to the centrifugal force effect and the poloidal centrifugal force effect can be neglected with $|\mathrm{v}_{\theta,z}/\mathrm{v}_{\varphi,z}| \sim O(\varepsilon)$; while in the edge regime with $|\mathrm{v}_{\theta,z}/\mathrm{v}_{\varphi,z}| \sim 1$, both the effects of the Coriolis force and the poloidal centrifugal force can not be neglected. Even though the important role of the poloidal centrifugal force on the in-out impurity density asymmetry that has been pointed out in [29, 30], the effect of the Coriolis force is much larger (factor $2q/\delta$) than that of the poloidal centrifugal force with $|v_{\theta,z}/v_{\varphi,z}| \sim 1$. Hence in the following, we will focus on the effect of the Coriolis force and neglect the poloidal centrifugal force.

Integrating Eq. (11), one finds

$$\frac{n_z(\theta)}{n_z(0)} = \left\{ -\frac{m_z}{T_z}\int_0^\theta (\omega_z^2 rR\sin\theta - 2qR_0\omega_z v_{\theta,z}\sin\theta)d\theta - \frac{Ze[\Phi(\theta)-\Phi(0)]}{T_z} \right\}. \quad (12)$$

To find $\Phi(\theta)-\Phi(0)$, we consider the deuterium as the bulk ion. The bulk ion density poloidal variation due to the bulk ion toroidal and poloidal rotation can be derived in a way similar to the above,

$$\frac{n_i(\theta)}{n_i(0)} = \exp\left\{ -\frac{m_i}{T_i}\int_0^\theta (\omega_i^2 rR\sin\theta - 2qR_0\omega_i v_{\theta,i}\sin\theta)d\theta - \frac{e[\Phi(\theta)-\Phi(0)]}{T_i} \right\}. \quad (13)$$

The inertia effect can be neglected for the electrons due to the small mass. Then we can have the Boltzmann relation $n_e(\theta)/n_e(0) = \exp\{e[\Phi(\theta)-\Phi(0)]/T_e\}$. For the trace impurities, $Zn_z \ll n_e$, the quasi-neutrality condition is simplified to $n_i(\theta) = n_e(\theta)$. Therefore we find

$$e[\Phi(\theta)-\Phi(0)] = -\frac{m_i T_e}{T_e+T_i}\int_0^\theta (\omega_i^2 rR\sin\theta - 2qR_0\omega_i v_{\theta,i}\sin\theta)d\theta. \quad (14)$$

This indicates that a poloidal electric field can be generated through the bulk ion poloidal redistribution due to the centrifugal force and the Coriolis force.

Substituting Eq. (14) into Eq. (12), we can obtain

$$\frac{n_z(\theta)}{n_z(0)} = \exp\left[ \frac{m_z rR_0\omega_z^2}{T_z}(1-\alpha_\varphi^2 \frac{Zm_i}{m_z}\frac{T_e}{T_e+T_i})(\cos\theta-1) \right.$$
$$\left. + \frac{2m_z qR_0\omega_z}{T_z}\int_0^\theta (1-\alpha_\varphi\alpha_\theta \frac{Zm_i}{m_z}\frac{T_e}{T_e+T_i}) v_{\theta,z}\sin\theta\, d\theta \right], \quad (15)$$

with $\alpha_\varphi = \omega_i/\omega_z$, $\alpha_\theta = v_{\theta,i}/v_{\theta,z}$. To obtain Eq. (15), the terms higher than $O(\varepsilon)$ have been neglected. For the DIII-D experiments [41], in which both the toroidal rotation of the impurities and the bulk ions are measured, $\alpha_\varphi \sim 1$ is estimated. We will assume $\alpha_\varphi \approx 1$ in the following. The poloidal rotation of the impurity and the bulk ion could be in the same or opposite direction, while $O(|\alpha_\theta|) \sim 1$ can be found from the experimental measurements [42].

The in-out asymmetry factor of the impurity density can be derived analytically with $v_{\theta,z(i)}(\theta) = \overline{v}_{\theta,z(i)}(1 + \delta_{z(i)} \cos\theta)$,

$$\frac{n_z(\pi)}{n_z(0)} = \exp\left\{-\frac{2m_z r R_0 \omega_z^2}{T_z}(1 - \frac{Zm_i}{m_z}\frac{T_e}{T_e + T_i}) + \frac{4m_z q R_0 \omega_z \overline{v}_{\theta,z}}{T_z}(1 - \overline{\alpha}_\theta \frac{Zm_i}{m_z}\frac{T_e}{T_e + T_i})\right\}, \quad (16)$$

where $\overline{\alpha}_\theta = \overline{v}_{\theta,i}/\overline{v}_{\theta,z}$. Eq. (16) is the main result of this Letter. In the bracket on the right-hand-side of Eq. (16), the first term and the second term are the effects of the centrifugal force and the Coriolis force, respectively, which are modified by the poloidal electric field through the terms proportional to Z. The effect of the centrifugal force pushes the impurities to the outboard side, this effect can be weakened by the poloidal electric field, as is similar to Refs. [17, 18].

The Coriolis force term can be negative or positive, depending on the sign of $q\omega_z \overline{v}_{\theta,z}$ when the effect of the poloidal electric field is ignored; note that the sign of $q$ represents the direction of screwing of the magnetic field line from the outboard side to the inboard side, as is illustrated in Fig. 1; the sign of $\omega_z \overline{v}_{\theta,z}$ represents the direction of the Coriolis force, as is also illustrated in Fig. 1. When the Coriolis force term has the same sign as that of the centrifugal force term, i.e., $q\omega_z \overline{v}_{\theta,z} < 0$, the out-in impurity asymmetry induced by the centrifugal force is further enhanced by the Coriolis force. When the Coriolis force term has the sign opposite to that of the centrifugal force term, i.e., $q\omega_z \bar{v}_{\theta,z} > 0$, the in-out impurity density asymmetry can be induced by the Coriolis force, if it is strong enough to overcome the centrifugal force effect.

It can be straightforwardly proved that the sign of $q\omega_z v_{\theta,z}$ can be simply determined in the following way: for $v_{\varphi,z}$ in the co-current direction, $q\omega_z v_{\theta,z}$ will be negative with $v_{\theta,z}$ in the ion-diamagnetic drift direction and positive with $v_{\theta,z}$ in the electron-diamagnetic drift direction; for $v_{\varphi,z}$ in the counter-current direction, $q\omega_z v_{\theta,z}$ will be negative with $v_{\theta,z}$ in the electron-diamagnetic drift direction and positive with $v_{\theta,z}$ in the ion-diamagnetic drift direction. This statement is also true for determining the sign of $q\omega_i v_{\theta,i}$.

It should be pointed out that the effect of Coriolis force can be modified by the poloidal electric field; when $f_{PE} \equiv 1 - \bar{\alpha}_\theta \dfrac{Zm_i}{m_z} \dfrac{T_e}{T_e + T_i} > 1$, the effect of the Coriolis force is enhanced by the poloidal electric field; when $0 < f_{PE} < 1$, the effect of the Coriolis force is weakened; when $f_{PE} < 0$, the effect of the Coriolis force is reversed. Note that the effect of poloidal electric field pointed out here is different from Refs. [25, 34].

3. Numerical example

As an application of the proposed theoretical model, we calculate the in-out density asymmetry of the impurity Boron $B^{5+}$ observed in a rotating ASDEX Upgrade tokamak plasma [29, 30]. The main parameters are $R_0 = 1.7m$, $a = 0.5m$, $B_{\varphi 0} = -2.5T$, $r/a = 0.995$, $q = -4.0$, $\omega_z = \omega_i = 6krad/s$, $\bar{v}_{\theta,z} = -10km/s$,

$T_z = T_i = 0.2 keV$. $\hat{e}_\varphi$ we have chosen is in the direction of the plasma current; the plasma current is in the counter-clockwise direction and the toroidal magnetic field is in the clockwise direction viewed from above; the toroidal rotation of $B^{5+}$ and $D^{1+}$ are both in the co-current direction; the impurity poloidal rotation is in the electron-diamagnetic drift direction. The sign convention is consistent with that in [29, 30]. Since there is no $v_{\theta,i}$ measurements, we assume $|\bar{\alpha}_\theta| \sim 1$ according to Ref. [42]. The in-out asymmetry factor of $B^{5+}$ is evaluated, according to Eq. (16), to be $n_z(\pi)/n_z(0) = 2.5, 1.64, 3.8$ for $\bar{\alpha}_\theta = 0.0, 1.0, -1.0$, respectively, which is qualitatively consistent with the experimental observations [29, 30]. The accurate quantitatively validations of the proposed model should be carried out with the measurements of both the poloidal rotation of the impurity and bulk ions in the tokamak experiments in the future.

The effects of the Coriolis force and the poloidal electric field on the in/out $B^{5+}$ density asymmetries are evaluated by scanning $\bar{v}_{\theta,z}$ and $\bar{\alpha}_\theta$. The contour plot of the in-out asymmetry factor of $B^{5+}$ in $\bar{v}_{\theta,z} - \bar{\alpha}_\theta$ plane is shown in Fig. 2. The bottom-left corner indicates an in-out asymmetry due to the Coriolis force (enhanced or weakened by the poloidal electric field); the bottom right corner indicates an out-in asymmetry due to the centrifugal force (enhanced or weakened by the Coriolis force and the poloidal electric field); the upper-right corner indicates an in-out asymmetry due to the effect of the Coriolis force reversed by the poloidal electric field; the upper-left corner indicates an out-in asymmetry due to the effect of the Coriolis force reversed by the poloidal electric field in addition to the centrifugal force. Therefore the in-out impurity asymmetry observed in the ASDEX Upgrade tokamak corresponds to the bottom-left corner, as is shown in Fig. 2; one concludes that the Coriolis force plays an important role in the in-out impurity asymmetry observed in this experiment.

4. Summary and conclusion

In summary, the effect of the Coriolis force due to the impurity toroidal and poloidal rotation on the in-out impurity density asymmetry in a rotating tokamak plasma is identified. The in-out impurity density asymmetry can be induced by the Coriolis force, if one sees a positive Coriolis force when moving along the magnetic field line from the outboard side to the inboard side in a magnetic flux surface. The proposed theoretical model is consistent with the in-out asymmetry of impurity density observed in the ASDEX Upgrade experiments [29, 30]. The in-out impurity asymmetry driven by the Coriolis force discovered here provides an opportunity for active controlling of the direction of the impurity neoclassical convection in the tokamak plasmas by controlling the plasma rotation.

**Acknowledgments**

This work was supported by the National Natural Science Foundation of China under Grant Nos. 11575246, 11375196, 11405174, 11505240 and the National Magnetic Confinement Fusion Program of China under Contract Nos. 2013GB107004, 2014GB113000.

**List of Figure Captions**

Figure 1. Illustration of the Coriolis force in the toroidally rotating frame in a tokamak plasma with $\omega_z \text{v}_{\theta,z} < 0$ for (a) $q < 0$; (b) $q > 0$.

Figure 2. Contour plot of the in-out density asymmetry factor of the impurity Boron $B^{5+}$ in $\bar{\text{v}}_{\theta,z} - \bar{\alpha}_\theta$ plane with $\omega_z = \omega_i = 6krad/s$, $q = -4.0$, $T_z = T_i = T_e = 0.2keV$. $\bar{\text{v}}_{\theta,z} > 0(<0)$ : the impurity poloidally rotates in the ion (electron)-diamagnetic drift direction. The color symbols represents $\bar{\text{v}}_{\theta,z} = -10km/s$ and $\bar{\alpha}_\theta = 1, 0, -1$, respectively.

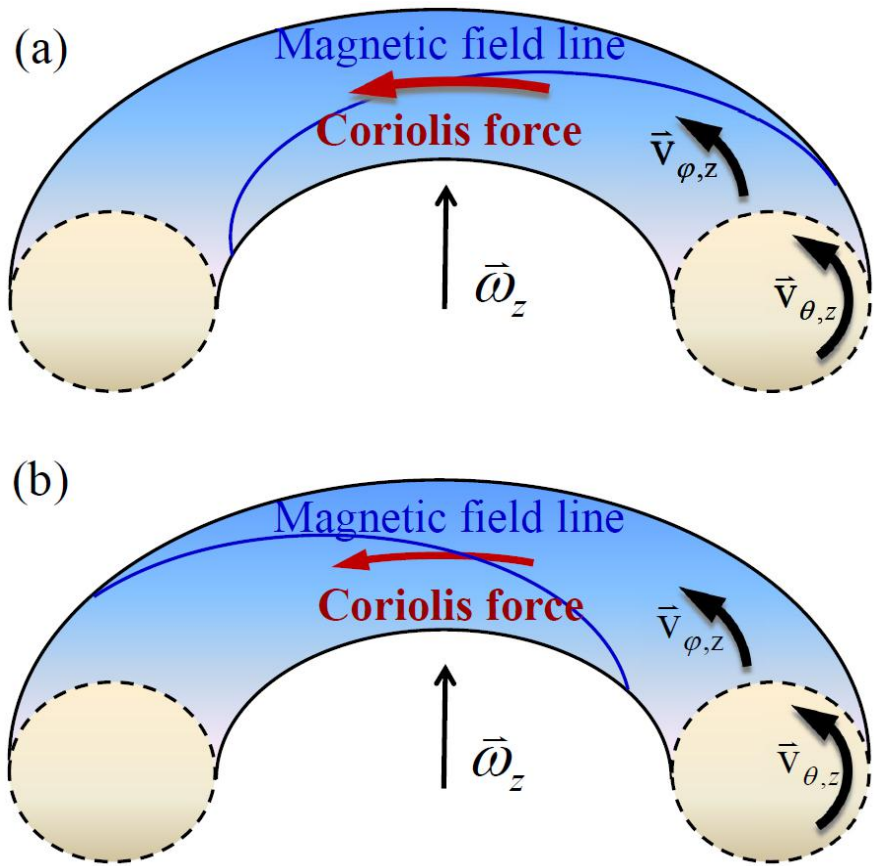

Figure 1.

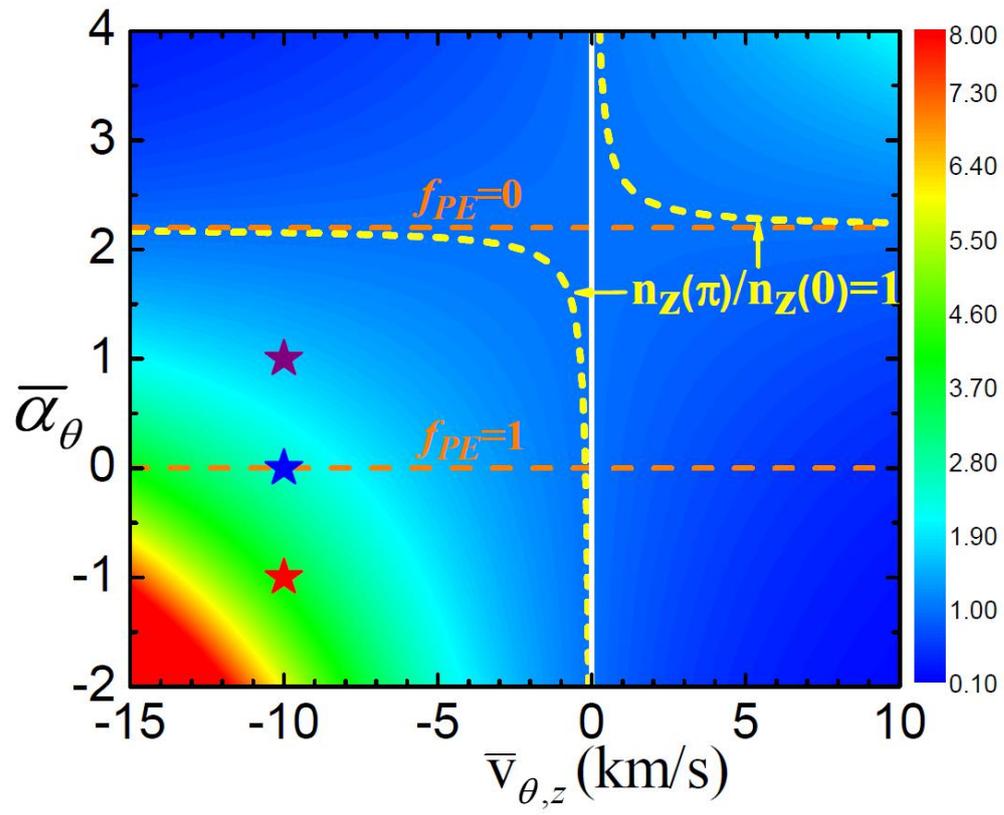

Figure 2.